\documentclass[pra,aps,twocolumn,showpacs]{revtex4} 
\newcommand{\pd }[1]{\frac \partial {\partial #1}} 
\newcommand{\pw }{\frac {\kappa ^2}{\Omega _\mathrm{C}^2} } 
\newcommand{\pddd }[2]{\frac {\partial #2}{\partial #1}}
\newcommand{\pvvv }[2]{\frac {\delta #2}{\delta #1}} 
\usepackage{epsfig}
\usepackage{color}

\begin{document} 
\title{Many-body physics of slow light} 
\author{I. E. Mazets} 
\affiliation{Wolfgang Pauli Institute c/o University of Vienna, 1090 Vienna, Austria;\\
Vienna Center for Quantum Science and Technology, Atominstitut, TU Wien, 1020 Vienna, Austria;\\ 
Ioffe Physico-Technical Institute, 194021 St.Petersburg, Russia} 

\begin{abstract} 
We present a quantum theory of slow light beyond the weak probe pulse approximation. By reduction of the full Hamiltonian of the 
system to an effective Hamiltonian for a single quantum field we demonstrate that the concept of dark-state polaritons can be 
introduced even if the linearized approach is no longer valid. The developed approach  allows us to study the evolution of non-classical 
quantum states of the polariton field.   
\end{abstract} 

\maketitle  

\section{Introduction} 
\label{secI} 

\begin{figure}[b]
\vspace*{2mm} 
\begin{center}
\epsfig{file=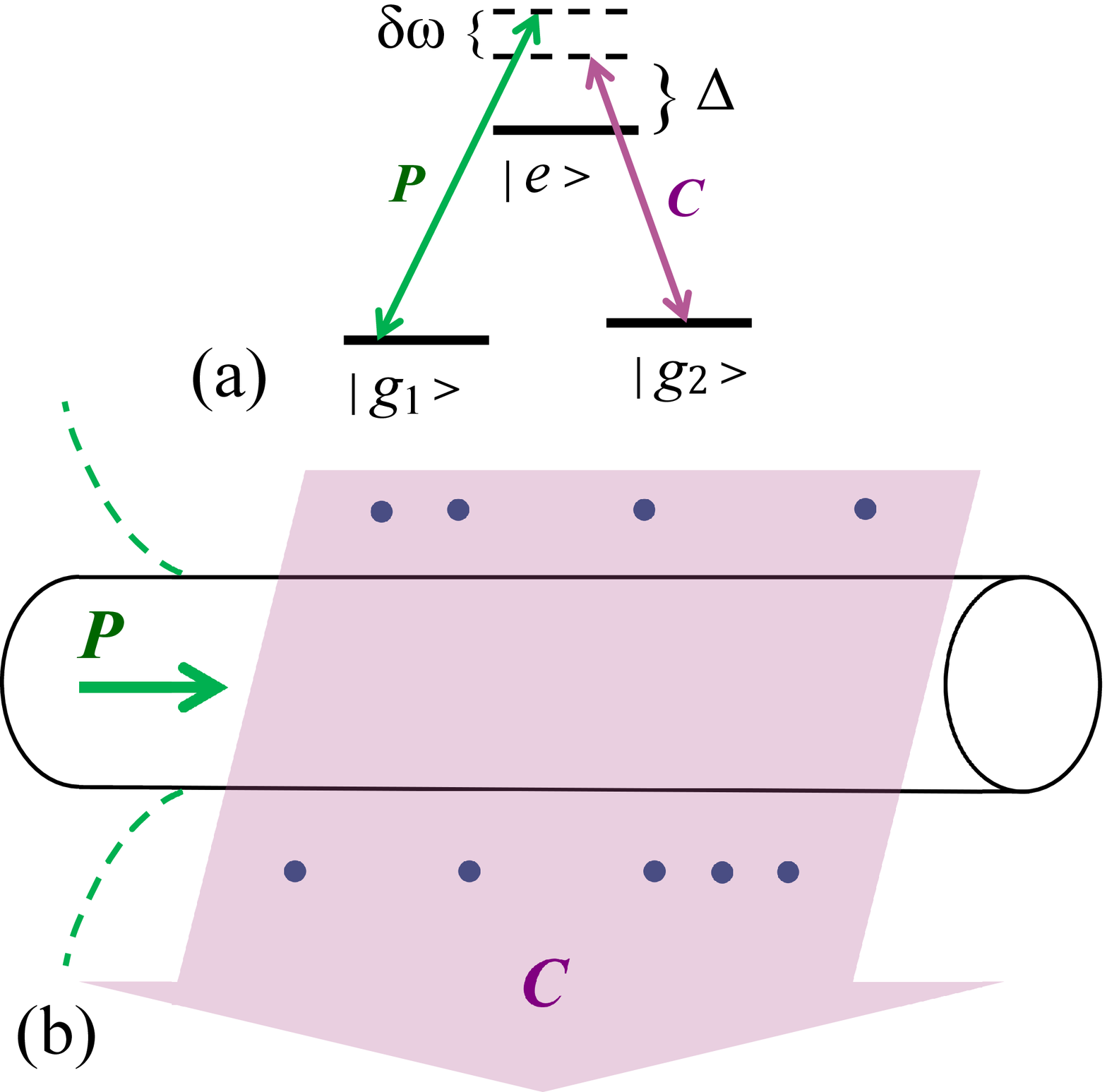,width=0.95\columnwidth } 
\end{center} 
\begin{caption}
{(Color Online). (a) The three-level excitation scheme of a slow-ligh esperiment: 
$|g_1\rangle $ and $|g_2 \rangle $ are ground state sublevels, $|e\rangle $ is an optically 
excited state. (b) Sketch of the atoms trapped near a nanofiber. The probe light is transmitted through the 
nanofiber (the dashed line shows its evanescent field that excites atoms). The coupling field is sent perpendicularly to the nanofiber. 
\label{F1} }
\end{caption}
\end{figure}

Slow light is a phenomenon associated with a propagation of dark-state polaritons \cite{Lukin1,Lukin2}, i.e.,  
quantum superpositions of photons and spin excitations in a three-level medium [see Fig.~\ref{F1}\, (a)], where a low-frequency 
coherence is established 
\cite{Harris1}, at the group velocity by many orders of magnitude less than the speed of light. 
The slow group velocity is attained via steep dispersion of the refractive index of the medium within the 
slow-light propagation window. Slow propagation of light pulses 
increases their interaction time and is therefore a key component of various proposed nonlinear-optical schemes aimed at the operation at the few-photon level 
\cite{propfew}. Adiabatic switching off the coupling field maps the photon state onto the collective spin state \cite{Lukin4}, 
thus providing a reversible quantum memory. 

Up to now the quantum aspects of the slow light propagation have been analyzed in the approximation, where the number of excitations in a medium is much less than the number of atoms \cite{Lukin1,Lukin2}. This approximation makes 
the construction of bosonic operators for dark-state polaritons easy and certainly holds for the case of 
experiments with atomic vapors in a gas cell \cite{expcell} or 
large ensembles of cold atoms \cite{expMOT}. Attempts to develop a quantum theory of 
slow light beyond the framework of Refs. \cite{Lukin1,Lukin2} have been scarce up to now \cite{beyond1,beyond2soliton} and the analysis of the models has proved itself to be  
quite difficult. Remarkably, the approach of Ref. \cite{Kuang} yielded definite results on the slow-light dynamics only in the 
semiclassical limit, and left open the question about the medium response to non-classical fields. 

It is intuitively clear  that, 
when the number of photons entering the medium becomes comparable to the number of atoms interacting with the light, the 
picture of dark-state polaritons with bosonic properties needs a more elaborate justification. 
The probe field interacts with a depleted medium and propagates at a velocity 
approaching the speed of light as the ratio of the input photons to the number of atoms increases. And a system that makes this situation experimentally feasible 
is now available. Thousands \cite{AR1} or hundreds (or tens) \cite{Kimble1} of atoms can be trapped near a single-mode tapered optical nanofiber and coupled 
via evanescent field to a probe (\textit{P}) radiation sent through the nanofiber. A similar physical situation can be achieved also for atoms in 
hollow-core fibers \cite{hcf}. 

To simplify the analysis, we assume that the classical coupling (\textit{C}) field is sent perpendicularly to the nanofiber, as shown in Fig.~\ref{F1}\, (b). 
This allows us to assume the Rabi frequency $\Omega _\mathrm{C}$ for 
the atomic transition driven by the coupling field to be constant along the nanofiber. The 
case of co-propagating, nanofiber-guided, quantized probe and coupling fields requires a more elaborate treatment and is beyond the scope of the present paper. 

The purpose of the present paper is to establish a many-body quantum theory of slow light for arbitrary probe light intensity, in other words, 
to formulate the problem as an effective Hamiltonian problem for a single bosonic field of dark-state polaritons beyond the linearized approach 
of Refs. \cite{Lukin1,Lukin2}. Although atoms in the system of interest do not interact with each other via short-range forces, nevertheless, they 
interact with each other via the nanofiber-guided electromagnetic  
field. Also we can say that  probe-field photons interact with each other via their coupling to the atomic medium. 
As a result, the collective field of dark-state polaritons emerges, in some analogy to collective excitations in Bose-Einstein condensates 
of weakly-interacting atoms \cite{PethickSmith}. This allows us to regard our theory as a many-body theory. 

In Sec. \ref{secII} we introduce the full Hamiltonian of the problem and recover, by its diagonalization in the single-excitation case, the weak-field 
limit \cite{Lukin1,Lukin2} for the group velocity of dark-state polaritons. In Sec. \ref{secIII} we recall the mean-field limit and the result of 
\cite{Kuang}. The derivation of the effective Hamiltonian for dark-state polaritons beyond the linearized (weak-field) approach is presented in 
Sec. \ref{secIV}. The analysis of the dynamics of non-classical states of the dark-state polariton field described by this effective Hamiltonian 
is the subject of Sec. \ref{secV}. 
 
\section{The full Hamiltonian}
\label{secII} 

We consider a one-dimensional (1D) system of $N$ three-level atoms, $|g_1\rangle $ and $|g_2\rangle $ being the ground state sublevels and $|e\rangle $ being an 
optically excited state. The classical coupling field is detuned from the $|g_2\rangle \leftrightarrow |e\rangle $ transition by the frequency $\Delta $ (we explicitly write this detuning for the sake of generality; however, our approach works 
also in the case $\Delta =0$). If the 
probe field  has the frequency $\omega $ then the two-photon detuning is $\delta   \omega = 
\omega -\omega _{eg_1} -\Delta $, where $\omega _{eg_1}$ is the resonant frequency  of the $|g_1\rangle \leftrightarrow |e\rangle $ transition, driven by 
the probe field. We use the slowly-varying amplitude approximation \cite{QO} for the probe field, choosing 
$\omega _0 =\omega _{eg_1} +\Delta $ as the carrier frequency. We assume the linear dispersion for the probe photons, 
$\omega =u|k|$, where $k$ is the wave number and $u$ is 
the velocity of  propagation in the nanofiber. The operator of annihilation of a photon in a 1D nanofiber at the point $z$ at time $t$ is 
$\hat {\cal E} (z,t)\exp[ -i \omega _0(t-z/u)]$. We introduce in a similar way the slowly varying amplitudes $\hat \psi _{1,2,e}$ for operators 
annihilating bosonic atoms in the states $|g_1\rangle $, $|g_2\rangle $, $|g_e\rangle $, respectively. Then  the full Hamiltonian 
of the system in the interaction representation and in the rotating wave approximation reads as  
\begin{eqnarray} 
\hat {H} &=&\hbar \int _0^L dz\, \Big{[} -iu\hat {\cal E} ^\dag \pd {z} \hat {\cal E} -\Delta \hat \psi _{e}^\dag \hat \psi _{e}- \kappa 
(\hat {\cal E} ^\dag \hat \psi _{1}^\dag \hat \psi _{e}+  \nonumber \\ && 
\hat \psi _{e}^\dag \hat \psi _{1}\hat {\cal E}) -\Omega _\mathrm{C} (\hat \psi _{2}^\dag \hat \psi _{e}+\psi _{e}^\dag \hat \psi _2)\Big{]} .
\label{fullH} 
\end{eqnarray} 
The periodic boundary conditions over the distance $L$ are assumed. We also recall that $\delta \omega =u\delta   k $, 
where $\delta   k = k-\omega _0/u$, reduces to 
$ -iu \pd {z}$ when we write the slowly varying amplitude of the probe photonic field in the co-ordinate representation. 
The coupling between the probe field 
and the atoms is given by $\kappa = d_{eg_1} \sqrt{\omega _{0}/(2\hbar \varepsilon _0A)}$, where 
$d_{eg_1}$ is the projection of dipole moment of the atomic transition driven by the probe light 
to the unit vector of the probe field polarization, 
$\varepsilon _0$ is the dielectric permittivity of the vacuum in SI units, $A$ is the effective mode area determined by the structure of the evanescent field 
\cite{Aeff}.  

Both the atom-number 
\begin{equation} 
\hat N =\int _0^L dz \left( \hat \psi _1^\dag \hat \psi _1+\hat \psi _2^\dag \hat \psi _2+\hat \psi _e^\dag \hat \psi _e\right) 
\label{opN} 
\end{equation} 
and the excitation-number 
\begin{equation} 
\hat M =\int _0^L dz \left( \hat {\cal E}^\dag \hat {\cal E}+\hat \psi _2^\dag \hat \psi _2+\hat \psi _e^\dag \hat \psi _e\right) 
\label{opM} 
\end{equation} 
oprators commute with the Hamiltonian (\ref{fullH}). 

It is easy to show  not only Eq. (\ref{opN}) holds, but also the 
operator of the local linear density of atoms is the integral  of motion: 
\begin{equation} 
\pd{t} \left( \hat \psi _1^\dag \hat \psi _1+\hat \psi _2^\dag \hat \psi _2+\hat \psi _e^\dag \hat \psi _e\right) =0.
\label{dens.const.t} 
\end{equation}
Exact positions of atoms near the nanofiber are not essential for our treatment. Hence, we introduce atomic field operators 
using some kind of a coarse graining \cite{Lukin1,Lukin2} over length scales exceeding the mean interatomic separation in 1D. 
In what follows we assume that the linear density of atoms $n_\mathrm{1D}=N/L$ is not only continuous, but also 
spatially uniform, 
\begin{equation} 
\pd {z} n_\mathrm{1D} =0 .
\label{dndzeq0} 
\end{equation} 

The Hamiltonian (\ref{fullH}) can be easily diagonalized for $M=1$. However, it is more instructive 
to find the eigenvalue of Eq. (\ref{fullH}) corresponding to the energy of a single dark-state polariton 
perturbatively, provided that the two-photon detuning is small enough, $|\delta \omega | \ll 
W_\mathrm{sl}$, where $W_\mathrm{sl}$ is the width of the slow-light propagation spectral window, 
discussed in the Appendix \ref{apX}.  
If the two-photon detuning is exactly zero, then the atomic medium  is in the dark state $|\mathrm{DS}\rangle $ 
characterized by 
\begin{equation} 
\hat \psi _e|\mathrm{DS}\rangle =0, \quad \left( \kappa {\hat {\cal E}}\hat \psi _1 +\Omega _\mathrm{C} \hat \psi _2\right) |\mathrm{DS}\rangle =0.
\label{DS} 
\end{equation} 
For small deviations from the two-photon resonance the energy $\hbar \omega ^\mathrm{DS}_{\delta  k}$ of the dark-state polariton 
can be calculated in the first order of the erturbation theory as 
\begin{equation}  
\hbar \omega ^\mathrm{DS}_{\delta  k} =u\delta   k \langle \mathrm{DS}|  \hat {\tilde {\cal E}}_{\delta  k} ^\dag  
\hat {\tilde {\cal E}}_{\delta   k} |\mathrm{DS}\rangle ,  
\label{first-order} 
\end{equation} 
where 
$$
\hat {\tilde {\cal E}}_{\delta  k}=\frac 1{\sqrt{L}} \int _0^L dz \, \hat {{\cal E}}\exp(-i{\delta  k}z) . 
$$
Then the group velocity of a single dark-state polariton $v_\mathrm{gr}= \partial \omega ^\mathrm{DS}_{\delta k} /(\partial \delta k)$ 
\cite{BornWolf} yields the well-known weak-field limit \cite{Lukin1,Lukin2}: 
\begin{equation} 
v_\mathrm{gr}^\mathrm{(w)} = \frac {u\varrho }{1+\varrho } , 
\label{weaklimit} 
\end{equation} 
where $\varrho = {\Omega _\mathrm{C}^2 /( \kappa ^2 n_\mathrm{1D})}$. If $\varrho \ll 1$, 
then the group velocity of the pulse is significantly slowed down compared to $u$. 

\section{The mean-field limit} 
\label{secIII} 

Heisenberg equations of motion for the electromagnetic and atomic field operators can be easily derived from Eq. (\ref{fullH}) using bosonic 
commutation rules. Then we assume the semiclassical (mean-field) approximation and substitute the operators by classical complex fields thus obtaining 
the following set of evolution equations (the time derivative being denoted by a dot): 
\begin{eqnarray} 
\dot{{\cal E}} &=& -u \pd{z}{\cal E} +i\kappa \psi _1^*\psi _2,           \label{cfE} \\
\dot{\psi }_1  &=& i\kappa {\cal E}^*\psi _e,                             \label{cf1} \\
\dot{\psi }_e  &=& i\Delta \psi _e+i(\kappa {\cal E}\psi _1+\Omega _\mathrm{C} \psi _2) ,    \label{cfe} \\
\dot{\psi }_2  &=& i\Omega _\mathrm{C} \psi _e.                           \label{cf2}
\end{eqnarray}  
Since in the slow-light regime the population of the optically excited state is negligibly small, we set 
\begin{equation} 
\kappa {\cal E}\psi _1+\Omega _\mathrm{C} \psi _2 = 0          \label{EITc} 
\end{equation} 
[cf. Eq. (\ref{DS})]. Eq. (\ref{EITc}) and Eq. (\ref{cf2}) rewritten as $\psi _e =-i\Omega _\mathrm{C}^{-1} \dot{\psi }_2$ reduce the number of 
independent variables to two. For them we obtain 
\begin{eqnarray} 
\dot{{\cal E}} + \frac {\kappa ^2}{\Omega _\mathrm{C}^2} \left( \psi _1^*\psi _1\dot{{\cal E}}+\psi _1^* {\cal E}
\dot{\psi }_1 \right)  &=& -u \pd{z}{\cal E},             \label{twoE} \\
\dot{\psi }_1 + \frac {\kappa ^2}{\Omega _\mathrm{C}^2} \left( {\cal E}^*{\cal E}\dot{\psi }_1+{\cal E}^*\psi _1 
\dot{{\cal E}} \right) &=&0.                              \label{two1} 
\end{eqnarray} 
From Eq. (\ref{two1}) we obtain the conservation law for the atom number in the  case, where all excitations are
dark-state polaritons  ($\psi _e^*\psi _e $ is negligible): 
\begin{equation} 
\psi _1^*\psi _1 \left( 1+ \frac {\kappa ^2}{\Omega _\mathrm{C}^2}  {\cal E}^*{\cal E}\right) =n_\mathrm{1D} . 
\label{n1D-cons}
\end{equation} 
Substituting Eq. (\ref{n1D-cons}) and following from Eq. (\ref{two1}) expression 
$$ 
\dot{\psi }_1 =-\frac {\kappa ^2 {\cal E}^* \psi _1 \dot{\cal E}}{{\Omega _\mathrm{C}^2} +  {\kappa ^2} {\cal E}^*{\cal E}} 
$$
into Eq. (\ref{twoE}), we obtain 
\begin{equation} 
\dot {\cal E} =-u \frac {({\Omega _\mathrm{C}^2} +  {\kappa ^2} {\cal E}^*{\cal E})^2}{\Omega _\mathrm{C}^2\kappa ^2n_\mathrm{1D} +
({\Omega _\mathrm{C}^2} +  {\kappa ^2} {\cal E}^*{\cal E})^2}  \pd{z} {\cal E} . 
\label{EE}
\end{equation} 
Hence, we obtained the propagation equation with the intensity-dependent group velocity of Ref. \cite{Kuang}. 

The intensity dependence of the group velocity has been studied in different contexts \cite{nn}. 
It manifests itself in wave front sharpening and, ultimately, in wave breaking \cite{ZaslUFN1}. 

\section{Derivation of the effective Hamiltonian for dark-state polaritons}
\label{secIV} 

The mean-field Eqs (\ref{twoE}, \ref{two1}) will be the starting point of our further derivations. We will reformulate the corresponding problem 
first in a Lagrangian and then in a Hamiltonian way. The resulting classical Hamiltonian will be again quantized and the quantum field for 
dark-state polaritons will be introduced. Such a method based on reduction of an exact many-body quantum problem to a  
set of classical Hamilton equations for certain collective variables and subsequent quantization of these collective variables proved 
itself to be successful in many studies, from the well-known quantization of phonons in solids \cite{SolidSt} to the theory of 
macroscopic quantum tunneling of a Bose-Einstein condensate with attractive interactions \cite{MQT}. 

\subsection{Classical variables and the effective Hamiltonian}
\label{ssIV-1} 

In what follows we use the Lagrangian and Hamiltonian formalisms for continuous systems described in detail, e.g.,  
in Ref. \cite{Goldstein}. 
We introduce the four generalized co-ordinates $J,\, S,\, \Xi,\, Q$ as real classical fields dependent on $z$ and $t$ via 
\begin{equation} 
{\cal E} =\sqrt{J}\exp (-iS/\hbar ) ,\quad {\psi _1} =\sqrt{{\Xi }}\exp (-iQ/\hbar ) . 
\label{defgencd} 
\end{equation}  
Obviously, $J\geq 0$ and $\Xi \geq 0$. Planck's constant appears in Eq. (\ref{defgencd}) in anticipation of the 
quantization of the variables in the next Subsection. 
Then the two complex Eqs. (\ref{twoE}, \ref{two1}) are transformed into four real equations 
\begin{eqnarray} 
\dot S + \pw \Xi (\dot S +\dot Q) &=& -u \pd{z}S ,                               \label{fr.S} \\
\dot J + \pw (\Xi \dot J+J\dot \Xi )&=&-u \pd{z}J ,    \label{fr.I} \\
\dot Q + \pw J(\dot S +\dot Q) &=&0,                                      \label{fr.Q} \\
\dot{\Xi } + \pw (\Xi \dot J+J\dot \Xi )&=&0 .                      \label{fr.X} 
\end{eqnarray} 
The Lagrangian $\Lambda \equiv \int _0^L dz\, {\cal L}$ is constructed in such a way that the Lagrangian equations 
$$
\frac d{dt} \frac {\delta  {\Lambda }}{\delta \dot q} = \frac {\delta  {\Lambda }}{\delta  q}
$$
where $\delta /(\delta q)$ stands for variational derivative or, equivalently, 
\begin{equation} 
\frac d{dt} \frac {\partial {\cal L}}{\partial \dot q} = \frac {\partial {\cal L}}{\partial  q}
\label{Lnew} 
\end{equation} 
is satisfied for $q$ standing for $J,\, S,\, \Xi,\, Q$. This determines the Lagrangian density 
\begin{equation} 
{\cal L} = u J\pd{z} S + \left( J+ \pw J\Xi \right) \dot S+ 
\left( {\Xi }+ \pw J\Xi \right) \dot Q .
\label{Lagrd}  
\end{equation}  
Note, that, due to the assumed periodic boundary conditions, integration by parts gives the result
\begin{equation} 
\int _0^L dz\, J\pd{z} S= -\int _0^L dz\, S\pd{z} J,      \label{ibp} 
\end{equation} 
which is used in derivation of Eq. (\ref{fr.I}) from Eq. (\ref{Lnew}). 

Since only $\dot S$ and $ \dot Q$, but not $\dot J$ and $ \dot \Xi $ appear in Eq. (\ref{Lagrd}), we introduce 
two generalized momenta 
\begin{eqnarray}
P_S &=& \pd{\dot{S}}{\cal L} =J+\pw J\Xi , \label{PS} \\ 
P_Q &=& \pd{\dot{Q}}{\cal L} =\Xi +\pw J\Xi . \label{PQ}
\end{eqnarray} 
Eq. (\ref{fr.X}) then reduces to $\dot P_Q =0$, where $P_Q$ has the meaning of the linear density of atoms in the mean-field limit 
under the slow-light propagation  conditions, i.e., $P_Q=n_\mathrm{1D}$. 

From Eq. (\ref{EITc}) we understand that $P_S$ is the sum of the densities of the photons and atoms driven from the state 
$|g_1\rangle $ to $|g_2\rangle $ under the slow-light 
propagation  conditions (without populating the optically excited state), i.e., the 
density of dark polaritons in the mean-field regime. Bright polaritons \cite{Lukin2}, 
excitations appearing when the condition (\ref{EITc}) is not fulfilled,  do not contribute to the value of $P_S$. 

Since $n_\mathrm{1D}$ is assumed to be spatially uniform [see 
Eq. (\ref{dndzeq0})], the introduction of the effective Hamiltonian 
\begin{eqnarray} 
H_\mathrm{eff}&=&  \int _0^L dz\, ( P_S\dot S+P_Q\dot Q-{\cal L}) \nonumber \\ 
&=& -u \int _0^L dz\, J(P_S,P_Q) \pd{z}S  
\label{Heff.cl} 
\end{eqnarray} 
easily yields the set of equations for the canonical variables 
\begin{eqnarray} 
\dot S &=&\pvvv{P_S}{H_\mathrm{eff}}=-u\pddd{P_S}{J(P_S,P_Q)}\pddd{z}{S} ,       \label{c.S}\\
\dot Q &=&\pvvv{P_Q}{H_\mathrm{eff}}=-u\pddd{P_Q}{J(P_S,P_Q)}\pddd{z}{Q} ,       \label{c.Q}\\
\dot P_S&=&-\pvvv{S}{H_\mathrm{eff}}=-u\pddd{P_S}{J(P_S,P_Q)}\pddd{z}{P_S} ,     \label{c.PS}\\
\dot P_Q&=&-\pvvv{Q}{H_\mathrm{eff}}=0,                                          \label{c.PQ} 
\end{eqnarray} 
which is equivalent to Eqs. (\ref{fr.S}\, --\, \ref{fr.X}). Taking the non-negative solution of  Eqs. (\ref{PS}, \ref{PQ}), 
we obtain 
\begin{eqnarray} 
J(P_S,P_Q)&=& \frac 12 \left( P_S -P_Q -\frac {\Omega _\mathrm{C}^2}{\kappa ^2}\right) + \nonumber \\ && 
\sqrt{  \frac 14 \left( P_S -P_Q -\frac {\Omega _\mathrm{C}^2}{\kappa ^2}\right)  ^2 + 
\frac {\Omega _\mathrm{C}^2}{\kappa ^2} P_S} \,  . ~ \, \, \, \, \, \, 
\label{JofPs} 
\end{eqnarray} 
After some algebra one can demonstrate the equivalence of Eqs. (\ref{c.S}, \, \ref{c.PS})  to Eq. (\ref{EE}). 

From now on, we consider $P_Q \equiv n_\mathrm{1D}$ as a mere constant and treat the Hamiltonian given by Eqs. 
(\ref{Heff.cl}, \ref{JofPs}) as a Hamiltonian for the canonical variables $S$ and $P_S$ only. 

By introducing the complex field 
\begin{equation} 
\Psi =\sqrt{P_S}\exp (-i S/\hbar ) 
\label{Psi.def} 
\end{equation} 
we can rewrite Eq. (\ref{c.S}, \ref{c.PS}) as 
\begin{equation} 
\pd{t}\Psi +v_\mathrm{gr} \pd{z} \Psi =0, 
\label{propag.cl} 
\end{equation} 
where 
\begin{equation} 
v_\mathrm{gr}=\frac u2 \left[ 1+\frac {\frac {P_S}{n_\mathrm{1D}} -1 +\varrho }{\sqrt{ \left(  
\frac {P_S}{n_\mathrm{1D}} -1 +\varrho \right) ^2 
+4\varrho }}\right]  
\label{vgr.cl} 
\end{equation} 
has the meaning of the intensity-dependent propagation velocity (group velocity) \cite{ZaslUFN1,NLW} of dark-state 
polaritons and  $\varrho = \Omega _\mathrm{C}^2/(\kappa n_\mathrm{1D})$ has to be much less than 1 to provide the 
slowdown of the propagation velocity of a  weak pulse. 
After some algebra Eq. (\ref{vgr.cl}) can be expressed in terms of the probe-field Rabi frequency 
$\Omega _\mathrm{P}=\kappa \sqrt{J}$, thus reproducing the result of Ref. \cite{Kuang}: $v_\mathrm{gr} = 
(\Omega _\mathrm{P}^2+\Omega _\mathrm{C}^2)^2 /[ (\Omega _\mathrm{P}^2+\Omega _\mathrm{C}^2)^2+\Omega _\mathrm{C}^2 
\kappa ^2n_\mathrm{1D}]$. The limit $P_S \rightarrow 0$ yields the well-known result Eq. (\ref{weaklimit}) in the 
weak-field limit \cite{Lukin1,Lukin2}.

\subsection{Quantization of the effective Hamiltonian} 
\label{ssIV-3}

In quantum theory, the canonic variables $S$ and $P_S$ can be replaced with 
operators $\hat S$ and $\hat P_S$  obeying  the commutation relation 
\begin{equation} 
[\hat S(z),\hat P_S(z^\prime )]=i\hbar \delta (z-z^\prime ) . 
\label{comm.bad} 
\end{equation} 
In principle, we could define a quantum field for dark-state polaritons using the analogy with the 
phase-density representation of the atomic field in the theory of degenerate gases of bosonic atoms \cite{MoraCastin}.  
Note that the sign of the commutator (\ref{comm.bad}) is opposite to the widely used convention  \cite{MoraCastin}, since it was 
natural to introduce $S$ in Subsec. \ref{ssIV-1} as a generalized co-ordinate; the standard definition of the phase and density 
operators implies choosing $P_S$ as a generalized co-ordinate and 
$-S$ as a generalized momentum. 
However, the phase-density representation is well defined on the length scales containing on average many field quanta. 
This is not a problem in theory of atomic Bose-Einstein condensates or quasicondensates \cite{MoraCastin}, but our goal is 
to formulate the theory in a way suitable for both small and large numbers of dark-state polaritons. 

We take therefore one more step in our classical treatment by transforming $S, \, P_S$ to new canonic 
variables 
\begin{equation} 
\psi _S =\sqrt{2\hbar P_S}\sin (S/\hbar ), \quad \psi _P =\sqrt{2\hbar P_S}\cos (S/\hbar ) ,
\label{new.cl}
\end{equation} 
${\cal F} = -\frac 12 \int _0^L dz\, \psi _S^2 \cot (S/\hbar )$ being the generating function of the canonical transformation. 
Obviously, $\Psi =({ \psi _P-i \psi _S})/{\sqrt{2\hbar }}$.  
And now we substitute the new canonic variables with the operators $\hat \psi _{S,P}$, which are Hermitian, since they 
correspond to the real-valued observables, and obey the canonical commutation rules 
\begin{eqnarray} 
{[\hat \psi _S(z),\hat \psi _S(z^\prime )]} &=& [\hat \psi _P(z),\hat \psi _P(z^\prime )]=0, 
\nonumber \\ 
{[\hat \psi _S(z),\hat \psi _P(z^\prime )]} &=& i\hbar \delta (z-z^\prime ). 
\label{comm.good}  
\end{eqnarray}
Then, in correspondence to Eq. (\ref{Psi.def}), we introduce the quantum field  
\begin{equation} 
\hat \Psi =\frac {\hat \psi _P-i\hat \psi _S}{\sqrt{2\hbar }}, \quad 
\hat \Psi ^\dag =\frac {\hat \psi _P+i\hat \psi _S}{\sqrt{2\hbar }} 
\label{Psi.q} 
\end{equation} 
that obeys the bosonic commutation relations 
\begin{eqnarray} 
{[\hat \Psi (z),\hat \Psi (z^\prime )]} &=&[\hat \Psi ^\dag (z),\hat \Psi ^\dag (z^\prime )]=0, 
\nonumber \\
{[\hat \Psi (z),\hat \Psi ^\dag (z^\prime )]} &=& \delta (z-z^\prime ). 
\label{comm.bos} 
\end{eqnarray} 
Recalling the physical meaning of $P_S$ as the semiclassical density of dark polaritons, we can identify $\hat \Psi (z)$ 
and $\hat \Psi ^\dag (z)$ with operators of annihilation and creation, respectively, of a dark-state polariton at the 
point $z$. The dark-state polariton density operator is, obviously, $\hat m _\mathrm{1D}(z) =\hat \Psi ^\dag (z)\hat \Psi (z)$, 
and $\hat M_\mathrm{D} =\int _0^L dz\, \hat m _\mathrm{1D}(z)$ is the operator of the total number of the dark-state 
polaritons with non-negative integer eigenvalues $M_\mathrm{D}$. 

Since the Hamiltonian (\ref{fullH}) reduces under the  conditions (\ref{DS}) to $\hat H_\mathrm{eff} =\hbar u \int _0^L dz\, 
\hat{\cal E}^\dag \left( -i \pd{z}\right) \hat{\cal E}$, we need to establish the relation between the probe field  
and dark-state polariton operators. A proper unitary transformation relates $\hat{\cal E}$ not only to $\hat \Psi $, but 
also to the field operator for bright-state polaritons \cite{Lukin2} and, in a general case, to the excitations of the type 
that gradually  approaches $\hat \psi _e$ as the atom-field coupling vanishes. But  dark state polaritons are decoupled from 
excitations of other types in the limit of adiabatically slow dynamics discussed in the Appendix \ref{apX}. 
Hence, we assume that only dark-state polaritonic excitations are present in the system, $M\equiv M_\mathrm{D}$, 
and relate $\hat{\cal E}$ to 
$\hat \Psi $. We assume this relation is local, i.e., contains only dark-state polariton density operator in a given point. 
The locality property helps us to infer this relation from an easily solvable case of $M_\mathrm{D}$ dark state polaritons 
created by coupling to the medium probe photons exactly at the two-photon resonance. We make our notation 
of the dark state more definite and 
explicitly write  the atom, $N$, and dark-state polariton, $M_\mathrm{D}$, quantum numbers. We introduce the annihilation 
operators $\hat a_q$ for probe photons and $\hat d_q$ for dark-state polaritons in the momentum modes via the plane wave 
expansions $\hat{\cal E}=\sum _q \hat a_q \exp (iqz)/\sqrt{L} $ and $\hat{\Psi }=\sum _q \hat d_q \exp (iqz)/\sqrt{L} $. Then 
the dark state $|\mathrm{DS};\, N, \, M_\mathrm{D}\rangle =(M_\mathrm{D}!)^{-1/2} d_0^{\dag \, M_\mathrm{D}}|0\rangle $, where 
$|0\rangle $ is the vacuum of polaritons, can be expressed, according to Eq.(\ref{DS}), as a superposition of products 
of Fock states of atoms in the states $|g_1\rangle $, $|g_2\rangle $ and of  probe  photons: 
\begin{eqnarray} 
|\mathrm{DS};\, N, \, M_\mathrm{D}\rangle &=& \frac 1{\sqrt{{\cal A}_{N,M_\mathrm{D}}} } 
\sum _{m=0}^{m_{\max }} 
(-1)^m \left( \frac \kappa {\Omega _\mathrm{C} \sqrt{L}}\right) ^m \times  \nonumber \\ && 
\sqrt { \frac {(-N)_m (-M_\mathrm{D})_m }{m!}   }  \times \nonumber \\ && 
|N-m\rangle _{g_1} |m\rangle _{g_2} |M_\mathrm{D}-m\rangle _\mathrm{phot}
, ~\label{darkF1} 
\end{eqnarray} 
where 
$$
(X)_m =\left \{ 
\begin{array}{ll} 
1, & m=0 \\
\prod _{j=1}^m (X+j-1), & m=1,\, 2,\, 3,\, \dots  
\end{array} 
\right. 
$$ 
is the Pochhammer symbol, $m_{\max } = {\min (N,M_\mathrm{D}) }$,  and the normalization factor is 
\begin{equation} 
{\cal A}_{N,M_\mathrm{D}} = \sum _{m=0}^{m_\mathrm{max} }\frac {(-N)_m (-M_\mathrm{D})_m }{m!} 
\left( \frac {\kappa ^2}{\Omega _\mathrm{C}^2{L}}\right) ^m 
.  \label{AM} 
\end{equation}

It is easy to show  that 
\begin{eqnarray} 
\hat a _0 |\mathrm{DS};\, N, \, M_\mathrm{D}\rangle &=& \sqrt{ {\cal Y}_{N,M_\mathrm{D}}  }
\sqrt{M_\mathrm{D}}|\mathrm{DS};\, N, \, M_\mathrm{D}-1\rangle \nonumber \\ &=& 
\sqrt{ {\cal Y}_{N,M_\mathrm{D}}  }\hat d_0|\mathrm{DS};\, N, \, M_\mathrm{D}\rangle , 
\label{darkF2} 
\end{eqnarray} 
where 
\begin{equation} 
{\cal Y}_{N,M_\mathrm{D}} =  \frac{{\cal A}_{N,M_\mathrm{D}-1}}{{\cal A}_{N,M_\mathrm{D}}} .
\label{YM} 
\end{equation} 
After some identical transformations we arrive at the following equation for ${\cal Y}_{N,M_\mathrm{D}}$: 
\begin{equation} 
{\cal Y}_{N,M_\mathrm{D}}=\frac {\kappa ^{-2}\Omega _\mathrm{C}^2L + ( {M_\mathrm{D}-1}) {\cal Y}_{N-1,M_\mathrm{D}-1} }{N+
\kappa ^{-2}\Omega _\mathrm{C}^2L + ( {M_\mathrm{D}-1}) {\cal Y}_{N-1,M_\mathrm{D}-1}}. 
\label{eq.Yex}
\end{equation} 
This exact equation can be used for recursive calculation of $ {\cal Y}_{N,M_\mathrm{D}}$ for increasing numbers of 
dark-state polaritons, starting from 
\begin{equation} 
{\cal Y}_{N,1}=\frac {\kappa ^{-2}\Omega _\mathrm{C}^2L  }{N   +\kappa ^{-2}\Omega _\mathrm{C}^2L }  .
\label{eq.Y1} 
\end{equation} 
On the other hand, we can make an assumption 
\begin{equation} 
{\cal Y}_{N,M_\mathrm{D}}\approx {\cal Y}_{N-1,M_\mathrm{D}-1}, 
\label{apY}
\end{equation}  
whose consisteny is easily checked \textit{a posteriori}. Then Eq. (\ref{eq.Yex}) reduces to a quadratic algebraic equation. 
Taking its positive root, we obtain
\begin{equation} 
{\cal Y}_{N,M_\mathrm{D}} \approx K \left( \frac {M_\mathrm{D}-1}L,\frac NL \right) , \quad 
K(P_S,P_Q)=\frac {J(P_S,P_Q)}{P_S} , 
\label{eq.Yqd} 
\end{equation} 
and the function $J(P_S,P_Q)$ is defined by Eq. (\ref{JofPs}). Note that Eq. (\ref{eq.Yqd}) reproduces Eq. (\ref{eq.Y1}) in 
the limit $M_\mathrm{D}=1$, where we take $K =\lim _{P_S \rightarrow 0} [ J(P_S,N/L)/P_S] =  
{\kappa ^{-2}\Omega _\mathrm{C}^2L  }/({N   +\kappa ^{-2}\Omega _\mathrm{C}^2L })$, recalling that all the variables 
in Eq. (\ref{JofPs}) are non-negative by definition.

\begin{figure}[t] 
\begin{center}
\epsfig{file=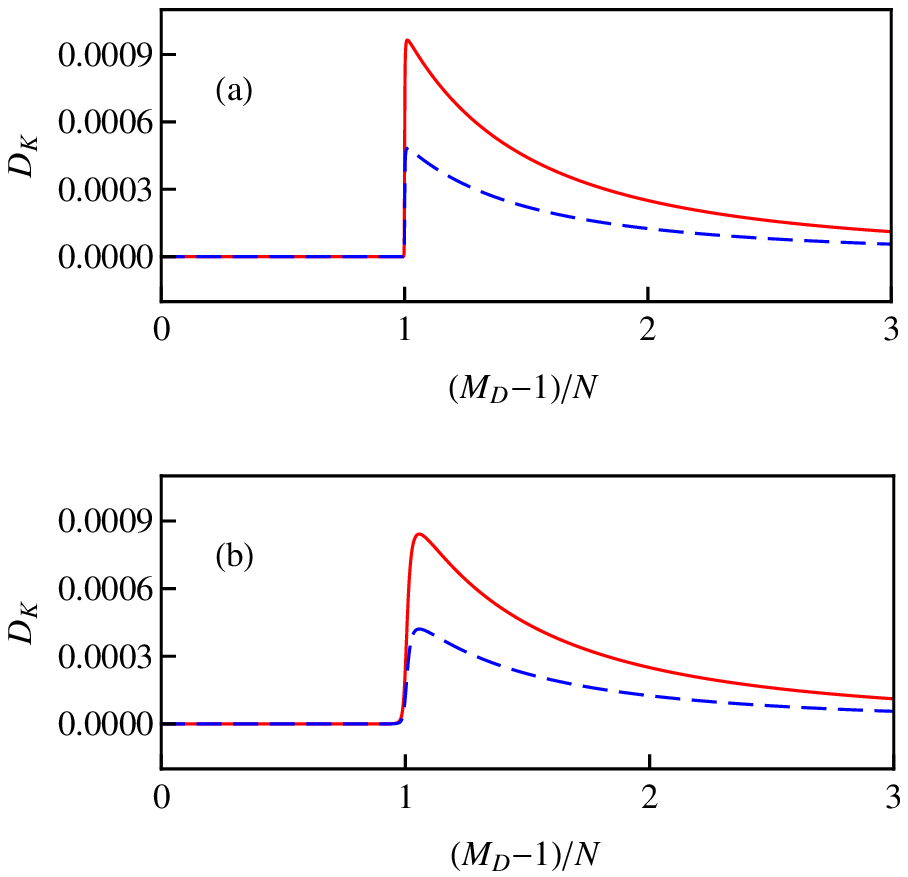,width=0.97\columnwidth } 
\end{center} 
\begin{caption}
{(Color Online). Difference $D_K = {\cal Y}_{N,M_\mathrm{D}} - K \left( \frac {M_\mathrm{D}-1}L,\frac NL \right)$  between the 
exact value of ${\cal Y}_{N,M_\mathrm{D}} $ defined by Eq. (\ref{YM}) and its approximation 
(\ref{eq.Yqd}) for $\varrho =10^{-6}$ (a) 
and $10^{-4}$ (b); $N=1000$ (solid line) and 2000 (dashed line). Units on the axes are dimensionless. }
\label{FF2}  
\end{caption}
\end{figure}

As we can see from Fig. \ref{FF2}, Eq. (\ref{apY}) provides a very good approximation for $M_\mathrm{D}-1<N$. 
The difference $D_K$ between the exact value of ${\cal Y}_{N,M_\mathrm{D}} $ and its approximation by Eq. (\ref{apY}) 
steeply rises near $(M_\mathrm{D}-1)/N=1$ to its maximum value $D_K^\mathrm{max} \sim 1/N$ and decreases slowly as 
$M_\mathrm{D}$ grows further. Such a deviation is, however, not important, since it is small 
compared to the limiting value of 1, which is rapidly approached by ${\cal Y}_{N,M_\mathrm{D}} $ as $\frac {M_\mathrm{D}-1}N$ 
begins to exceed unity by more than 
$2\sqrt{\varrho }$. For pulses of finite spatial extension $\ell _\mathrm{p}$ in the medium, we estimate the maximum 
systematic error of our approximation (\ref{eq.Yqd}) as $D_K^\mathrm{max}\sim 1/(\ell _\mathrm{p} n_\mathrm{1D})$, which is 
always much less than unity, since by our course-graining assumption there are many atoms on a typical length scale of the problem.   

Using Eq. (\ref{eq.Yqd}) we obtain 
\begin{equation} 
\hat{\cal E}\hat \varpi _\mathrm{D} = \sqrt{ K( \hat \Psi ^\dag \hat \Psi ,n_\mathrm{1D}) } \, \hat \Psi , 
\label{E.Psi} 
\end{equation} 
where $\hat \varpi _\mathrm{D} =\sum _{M_\mathrm{D}=0}^\infty |\mathrm{DS};\, N, \, M_\mathrm{D}\rangle 
\langle \mathrm{DS};\, N, \, M_\mathrm{D}| $ is the projection operator to the Hilbert subspace containing only 
dark-polariton states. Placing $\sqrt{K}$ to the \textit{left} from the dark-polariton annihilation operator enables us 
to substitute $(M_\mathrm{D}-1)/L$ by the operator of the local density of dark-state polaritons 
$\hat \Psi ^\dag \hat \Psi $ in the first argument of $K$. Finally, the quantum effective 
Hamiltonian for dark-state polaritons is 
\begin{eqnarray} 
\hat H_\mathrm{eff}&=& \hbar u \int _0^L dz\, \hat \Psi ^\dag \sqrt{ K( \hat \Psi ^\dag \hat \Psi ,n_\mathrm{1D}) }
\times \nonumber \\ &&  
\qquad \left( -i\pd{z}\right) \sqrt{ K( \hat \Psi ^\dag \hat \Psi ,n_\mathrm{1D}) }\, \hat \Psi . 
\label{Heff.neu} 
\end{eqnarray}  

\section{Results and discussion} 

\label{secV}

The most interesting application of the theory developed in the previous section concerns the dynamics of 
non-classical states, which is hardly accessible by the methods developed previously \cite{Kuang}. 
We can easily generalize the variational approach \cite{PethickSmith} to these states. 
We assume a probe state $| {\chi } \rangle $ characterized by certain variational parameters and find 
an extremum 
\begin{equation} 
\delta {\cal S}_\chi =0
\label{ext.s} 
\end{equation} 
of the action 
\begin{equation} 
{\cal S}_\chi = \langle \chi | \int _{t_1}^{t_2}dt \left(i\hbar \int _0^L dz\, \hat \Psi ^\dag \pd{t}\hat \Psi 
-\hat H_\mathrm{eff} \right) |\chi \rangle .   
\label{act}  
\end{equation}  

\begin{figure}[t] 
\begin{center}
\epsfig{file=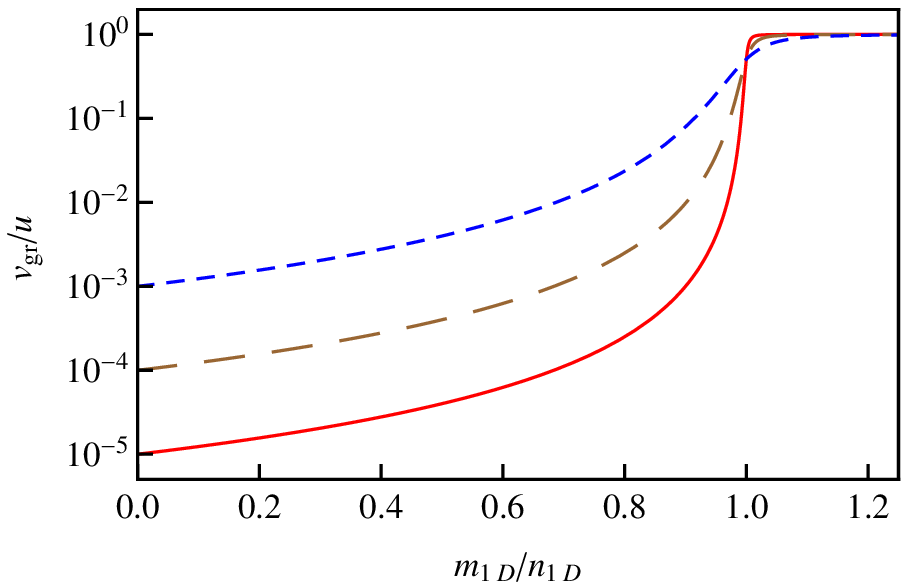,width=0.97\columnwidth } 
\end{center} 
\begin{caption}
{(Color Online). Group velocity of dark-state polaritons (normalized to the phase velocity $u$ of the probe light in the 
nanofiber) on the logarithmic scale as a function of the ratio of the dark-state polariton 1D density to the atomic 1D density. 
$\varrho =10^{-5}$ (solid line), $10^{-4}$ (long-dashed line), and $10^{-3}$ (short-dashed line). 
The units on the axes are dimensionless.     }\label{vgr2u}  
\end{caption}
\end{figure}

An exemplary non-classical state is a Fock state. Assume that $M_\mathrm{D}$ 
dark polaritons occupy the same state corresponding to 
a wave packet with a slowly varying envelope $\Phi (z,t)$ (the normalization $\int _0^L dz \, |\Phi |^2 = 1$ is assumed), 
i.e., $|\chi \rangle = (M_\mathrm{D}!) ^{-1/2} \hat d _\Phi ^{\dag \, M_D}|0\rangle $, where $|0\rangle $ is the polaritonic 
vacuum state and $\hat d_\Phi ^\dag $ creates a dark polariton in the wave-packet state. Then Eq. (\ref{ext.s}) reads 
explicitly as 
$$
\frac {\delta S_\chi }{\delta \Phi ^*}=0, 
$$
which results in the evolution equation  for the slowly-varying envelope 
\begin{equation} 
\pd{t}\Phi +v_\mathrm{gr} \pd{z} \Phi =0 , 
\label{Fock1} 
\end{equation} 
where $v_\mathrm{gr} $ is given by Eq. (\ref{vgr.cl}) with $P_S = (M_\mathrm{D}-1)|\Phi |^2 \equiv m_\mathrm{1D}$. 
Eq. (\ref{Fock1}) can be solved by the characteristics method \cite{NLW,char}. 

\begin{figure}[t] 
\begin{center}
\epsfig{file=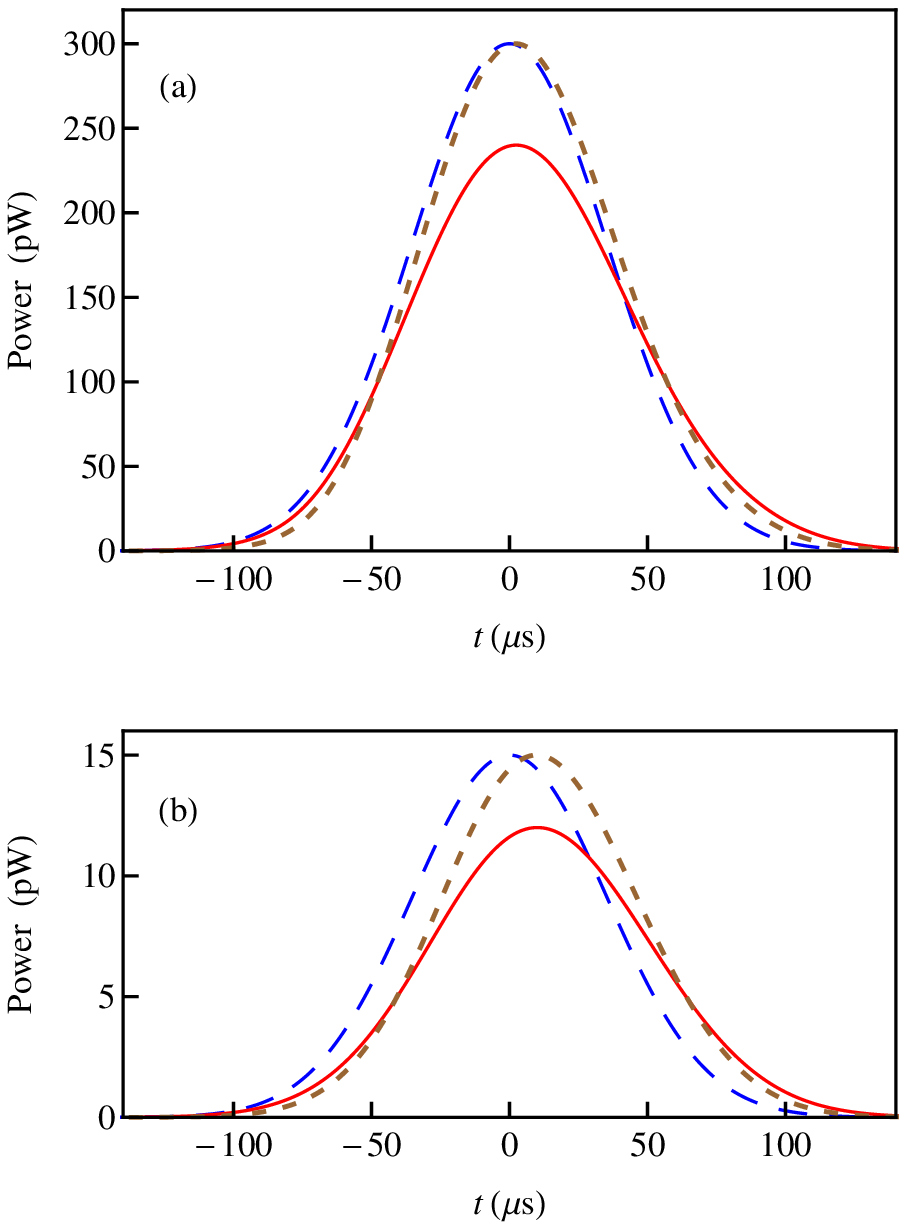,width=0.97\columnwidth } 
\end{center} 
\begin{caption}
{(Color Online). Power of the probe field at the entrance (long-dashed line) 
and at the exit (solid line) of the nanofiber-coupled atomic medium for a strong (a) and weak (b) pulse. 
See the system parameters in the text. Effects of absorption 
are taken into account according to Eq. (\ref{Fock1A}). 
As a guide for eye, we show by a short-dashed line the pulse at the exit of 
the medium with fully neglected absorption [Eq. (\ref{Heff.neu})]. 
In the case (a) the delay of the pulse peak arrival is negligible compared to the pulse peak delay of about 10~$\mu $s in the case (b).    }\label{Fold4} 
\end{caption}
\end{figure} 

Note that the propagation of the probe light intensity of the pulse is described by the essentially classical 
nonlinear group velocity even for a Fock state of dark-state polaritons, where $\langle \hat \Psi \rangle =0$. 
The number of probe photons is not well defined in this case, however, the state of the probe light is entangled with 
the state of the atomic medium [see Eq. (\ref{darkF1})], and the average amplitude of the probe light is also zero. 
The absence of such coherences, contrary to the concerns of Ref. \cite{Kuang}, does not change the dynamics dramatically, 
compared to the classical limit. This is not very surprising, since the optical coherence is shown to be a sufficient, 
but not necessary condition for observing many phenomena, traditionally associated with the semiclassical regime \cite{convf}. 

The group velocity shown in Fig. \ref{vgr2u} exhibits rapid saturation at $v_\mathrm{gr}=u$ for $m_\mathrm{1D}>n_\mathrm{1D}$. 
Such a behavior can be associated with the depletion of the state $|g_1\rangle $ at too a high density of the dark-state polaritons: 
although the system remains in the dark state defined by Eq. (\ref{DS}), the 
average number of atoms available for coupling to the probe light becomes small, and probe photons pass through the medium without 
interaction-induced delay. 

As an illustration, we present in Fig. \ref{Fold4} the results of numerical calculations of a slow-light pulse propagation through a nanofiber of a length $L=5000~\mu \mathrm{m}$. 
The linear density of cesium atoms \cite{Steck} coupled to the nanofiber is $n_\mathrm{1D}=1~\mu \mathrm{m}^{-1}$, the effective area of the probe 
field mode $A=3~\mu \mathrm{m}^2$.  
We assume that the probe field drives the $|g\rangle =|F=4,\, M_F=4\rangle \leftrightarrow |e\rangle =|F^\prime =5,\, 
M_F^\prime =3\rangle $ transition of the cesium $D_2$-line. The phase velocity $u$ of light in the nanofiber is assumed to be of about 0.9 speed of light in vacuum; 
$\Omega _\mathrm{C}=3\times 10^6~\mathrm{s}^{-1}$. 

Up to now, 
in our fully Hamiltonian theory we neglected the decay of the dark-state polaritons due to the small, but non-zero population of the 
optically excited state and coupling of the optical transitions to the free-space electromagetic modes. This assumption is 
valid if the two-photon detuning is less than the 
slow-light propagation spectral window, which is of the order of $\Omega _\mathrm{C}^2/(\gamma 
\sqrt{s})$ \cite{Harris1,Matisov1} for an optically dense ($s>1$) medium, 
where $2\gamma $ is the radiative decay rate of the optically excited state,  
$s=L/\zeta $ is the optical density of the medium, $\zeta = A/(n_\mathrm{1D}\sigma _0)$ is Beer's length, 
and $\sigma _0 $ is the resonance cross-section of the probe light absorption. 

Absorption effects can be accounted for by adding a corresponding non-adiabatic term \cite{Lukin2} to the propagation equation, which then 
reads as 
\begin{equation} 
\pd{t}\Psi +v_\mathrm{gr} \pd{z} \Psi = \frac {v_\mathrm{gr}^3\gamma ^2}{2\zeta \Omega _\mathrm{C}^4}
\frac {\partial ^2}{\partial z^2}\Psi ,  
\label{Fock1A} 
\end{equation} 
where by $\Psi $ we now denote the product of the normalized envelope function and the square root of the mean number of dark-state 
polaritons in the pulse. The expression $\Psi =\langle M_\mathrm{D}\rangle  ^{1/2} \Phi (z,t)$ is suitable in both the 
semiclassical and the Fock-state cases. The mean 1D density of probe photons  is then approximately 
$|\Psi |^2 K[(\langle M_\mathrm{D}\rangle -1)|\Phi |^2 ,n_\mathrm{1D}]$. 
For the parameters of Fig. \ref{Fold4} ($s\sim 250$) absorption begins to play a role, but does not destroy the pulse too much.  
The delay time of the pulse peak arrival remains the same, the pulse becomes slightly broadened because of preferential 
absorption of its high-frequency components. 

Generalizations of our variational theory  in the spirit of multiconfigurational variational method \cite{Alon}   
are possible, however, their development is out of the scope of the present  paper. 

To summarize, we developed a quantum many-body theory for the propagation of slow-light pulses. 
We developed a quantization framework that enabled us to introduce a bosonic quantum field for dark-state polaritons. 
The effective quantum Hamiltonian (\ref{Heff.neu}) is the main result of our work. 
We considered atoms coupled to a nanofiber as a definite example of an atomic medium, however, our results may be easily 
generalized to the cases of laser beam propagating in a gas cell or in an ultracold atomic cloud by replacing $A$ by the 
effective cross-section area of the probe beam. We found that the propagation of non-classical wave packets of slow light 
(Fock states of dark-state polaritons) is very similar to the classical dynamics in terms of light intensity. The existence of 
probe-field coherences is not necessary,  contrary to the expectations of Ref. \cite{Kuang}.

The author thanks A. Rauschenbeutel, C. Sayrin,  Ph. Schneewei{\ss}, and E. Shahmoon for helpful discussions. 
This work is supported by the Austrian Science Fund (FWF), project P 25329-N27. 

\appendix

\section{Spectral width of the slow-light propagation regime} 
\label{apX} 

The width $W_\mathrm{sl}$ of the slow-light propagation spectral window in optically dense ($s>1$) medium 
is well-known, see, e.g., 
the absorptive term in the bright-state polariton propagation equation in Ref. \cite{Lukin2}. Locally, the two-photon 
detuning couples the dark state to the bright state. The latter is coupled to the optically excited state and therefore has 
the width equal to the rate of induced transition to the optically excited state \cite{Matisov1}. If the two-photon 
detuning exceeds this width, the dark- and bright-states become mixed, and all effects based on the existence of the dark state 
decoupled from the optically excited state, including the slowing down of the probe pulse propagation, disappear. 
The effects of absorption in the medium further reduce this width by a factor of $1/\sqrt{s}$. 

In this Appendix we consider in detail the case of a very large one-photon detuning, i.e., we consider 
the Hamiltonian (\ref{fullH}) with $\Delta $ being the largest frequency available in the system; also $\Delta $ is 
assumed to be so large that the natural width of the optically excited state and the related effects of absorption of the 
probe photons can be neglected. We consider also, for the sake of clarity, states with a single excitation ($M=1$). 
We denote the states as follows: $|1\rangle $ is the state where all stoms are in their internal state $|g_1\rangle $ and one 
photon is present; in the state $|2\rangle $ there are no photons, but one atom is transferred from $|g_1\rangle $ to 
$|g_2\rangle $; the state with no photons and one atom excited to the state $|e\rangle $ is denoted by $|3\rangle $. 
Since the one-photon detuning is the highest frequency in this system, the state $|3\rangle $ can be adiabatically 
eliminated. For the probability amplitudes $a_j$, $j=1,2$, of the two remaining states we obtain the following 
Schr\"{o}dinger equation: 
\begin{equation} 
i\pd{t} \left( 
\begin{array}{l} 
a_1\\
a_2 \end{array} \right) = \left( 
\renewcommand{\arraystretch}{1.6}
\begin{array}{cc} 
\delta \omega +\frac {\kappa ^2 n_\mathrm{1D} }\Delta ~&  \frac {\kappa \sqrt{ n_\mathrm{1D}}\Omega _\mathrm{C}}\Delta \\  
\frac {\kappa \sqrt{ n_\mathrm{1D}}\Omega _\mathrm{C} }\Delta ~& \frac { \Omega _\mathrm{C}^2 } \Delta 
\end{array} \right)    \left( 
\begin{array}{l} 
a_1\\
a_2 \end{array} \right). 
\label{A.1}  
\end{equation} 
We analyze the eigenvalues and eigenvectors of Eq. (\ref{A.1}) depending on the two-photon detuning $\delta \omega = 
u\delta k$. The two states, denoted by superscripts $(\pm )$, and their respective eigenfrequencies are given by 
\begin{equation} 
\left( 
\begin{array}{l} 
a_1^{(+)}\\
a_2^{(+) }\end{array} \right) = \left( 
\begin{array}{l} 
\sin \vartheta \\
\cos \vartheta  \end{array} \right) , \quad 
\left( 
\begin{array}{l} 
a_1^{(-)}\\
a_2^{(-) }\end{array} \right) = \left( 
\begin{array}{r} 
\cos \vartheta  \\
-\sin \vartheta  \end{array} \right) ,
\label{A.2} 
\end{equation} 
\begin{eqnarray} 
\omega _{\delta k}^{(\pm )}& =&\frac 12 \left( \frac {\kappa ^2n_\mathrm{1D} +\Omega _\mathrm{C}^2}\Delta 
+\delta \omega \right) \nonumber \\ && 
\pm \sqrt{  \frac 14 \left( \frac {\kappa ^2n_\mathrm{1D} +\Omega _\mathrm{C}^2}\Delta 
+\delta \omega \right)^2 -\frac {\Omega _\mathrm{C}^2\delta \omega }\Delta  } \,  , ~~
\label{A.3} 
\end{eqnarray} 
where  
\begin{equation} 
\cot \vartheta = \frac {\Omega _\mathrm{C}}{\kappa \sqrt{n_\mathrm{1D}}} \left( 1- 
\frac {\delta \omega }{\omega _{\delta k}^{(+ )}}\right) . 
\label{A.4} 
\end{equation} 

The dark-polariton state admitting the slow light propagation satisfies two conditions: (i) the derivative of its eigenfrequency 
over $\delta k$, i.e., the group velocity of the excitatio, is small compared to $u$ and (ii) the state adiabatically reduces to 
$|1\rangle $ when the ratio $\kappa \sqrt{n_\mathrm{1D}} /\Omega _\mathrm{C}$ is 
formally decreased to 0 (i.e., the coupling between $|g_1\rangle $ and $|e\rangle $ 
is switched off). Eqs. (\ref{A.2}~---\ref{A.4}) show 
that the state $|(-)\rangle $ possesses these properties for $|\delta \omega |\ll W_{sl} = 
(  {\kappa ^2n_\mathrm{1D} +\Omega _\mathrm{C}^2}  )/|\Delta |$. Outside this spectral range either the group velocity is high 
(close to $u$) or the state does not reduce to $|1\rangle $ in the limit of the vanishing coupling between 
$|g_1\rangle $ and $|e\rangle $. In the latter case, a wave packet containing different photonic wave numbers and 
having adiabatically slowly changing envelop transforms, after entering the medium, into an excitation with the 
group velocity $\sim u$ with overwhelming probability. 

If the one-phonon detuning is not too large, $|\Delta |\lesssim \sqrt{\kappa ^2n_\mathrm{1D} +\Omega _\mathrm{C}^2}$, then the 
adiabaticity condition requires 
$|\delta \omega |\ll \sqrt{\kappa ^2n_\mathrm{1D} +\Omega _\mathrm{C}^2}$ \cite{Lukin2,propfew,nn}.

\end{document}